\documentstyle[preprint,aps]{revtex}
\tighten
\draft
\begin{document}
\widetext
\input epsf
\preprint{HUTP-98/A030, NUB 3177}
\bigskip
\bigskip
\title{Large $N$ Domain Walls as D-branes for ${\cal N}=1$ QCD String}
\medskip
\author{Gia Dvali$^{1}$\footnote{E-mail: 
dvali@ictp.trieste.it} and 
Zurab Kakushadze$^{2,3}$\footnote{E-mail: 
zurab@string.harvard.edu}}

\bigskip
\address{$^1$International Centre for Theoretical Physics, 34100 Trieste, Italy\\
$^2$Lyman Laboratory of Physics, Harvard University, Cambridge, 
MA 02138\\
$^3$Department of Physics, Northeastern University, Boston, MA 02115}
\date{July 19, 1998}
\bigskip
\medskip
\maketitle

\begin{abstract}
{}We consider a model which in a certain limit reduces to the large $N$ ${\cal N}=1$ 
supersymmetric $SU(N)$ gauge theory without matter. 
The gaugino condensate in this model is controlled by the dynamics of an additional singlet superfield. Using this model we explicitly construct
BPS domain walls arising due to the chiral symmetry breaking. In particular, in the large $N$ limit
we obtain the exact shapes of the domain walls corresponding to solitons, and also
of the domain walls interpreted as D-branes on which the SQCD string can end, whose existence was previously argued by Witten in the context of the large $N$ SQCD. We also discuss various points which appear to
support the consistency of the D-brane interpretation for these domain walls within the SQCD string context. 
\end{abstract}

\section{Introduction}

{}One of the longstanding problems in high energy physics is to gain analytical control over 
strongly coupled gauge theories, and, in particular, QCD. Many fascinating aspects of QCD, such as confinement, mass gap and chiral symmetry breaking, still lack desired quantitative
description, albeit over the years much qualitative (and numerical) insight has been gained into these phenomena. The closest supersymmetric relative of QCD, namely, pure ${\cal N}=1$ supersymmetric $SU(N)$ gauge theory {\em a priori} offers a possibility of simplifications (due to supersymmetry), 
yet it shares most of the qualitative features of QCD. Thus, SQCD without matter is believed to
be a confining theory with a mass gap and chiral symmetry breaking. It is therefore reasonable to suspect that a better understanding of pure SQCD might shed light on these phenomena in a more general context.

{}It is not unreasonable to believe that confinement and chiral symmetry breaking might be intimately related \cite{am}. 
If so, understanding chiral symmetry breaking in SQCD might shed some light
on the confining properties of (super)glue. Chiral symmetry breaking in SQCD is due to gaugino condensation. Thus, the composite operator $\lambda\lambda$ acquires a non-zero expectation value in strongly coupled SQCD: $\langle\lambda\lambda\rangle_k=a(N)\exp(2\pi ik/N) \Lambda^3$, where $\Lambda$ is the dynamically generated scale of SQCD, and $a(N)$ is an overall factor which depends on the subtraction scheme. There are $N$ inequivalent vacua
labeled by $k=0,\dots,N-1$. This implies that there should exist domain walls connecting pairs of distinct vacua. In \cite{wall} these domain walls were argued to be BPS saturated. The BPS property of the SQCD domain walls suggests that we might be able to gain some analytical
control over these solutions, and, therefore, over chiral symmetry breaking. 

{}A seemingly unrelated subject is 't Hooft's large $N$ limit \cite{thooft}. In \cite{thooft} it was noticed that in this limit the gauge theory diagrams organize themselves in terms of Riemann
surfaces characterized by the numbers of boundaries and handles, and addition of each extra handle on the surface corresponds to suppression by a factor of $1/N^2$. This observation has led 't Hooft to speculate that the large $N$ gauge theory might be described by some kind of string theory. Then the large $N$ expansion of gauge theories would be mapped to the string
expansion in terms of the properly weighted world-sheets of various topologies.

{}In the context of weakly coupled gauge theories ({\em i.e.}, when the effective
gauge coupling $\lambda=Ng^2_{YM}$ is fixed at a value $\lambda{\ \lower-1.2pt\vbox{\hbox{\rlap{$<$}\lower5pt\vbox{\hbox{$\sim$}}}}\ }1$) the first concrete realization of this idea was given in \cite{CS} in the context of three dimensional Chern-Simons gauge theory where the boundaries of the string world-sheet are ``topological'' 
D-branes. More recently, string expansion was shown to precisely reproduce 't Hooft's large $N$ expansion for certain four dimensional gauge theories \cite{BKV}\footnote{This was subsequently generalized to include unoriented world-sheets in \cite{ZK}.}. Here the boundaries are Type IIB D3-branes, and the (open) QCD string is simply the open string stretched between these D-branes.

{}As to the case of strongly coupled gauge theories, the story here appears to be much more complicated. In particular, it is unclear what is the string theory that governs the dynamics of
the QCD string which is expected to arise as an effective description in the large $N$ limit. This string theory is expected to be non-critical \cite{polyakov}, which makes it difficult to study. 
Nonetheless, in \cite{witten} Witten has made a remarkable observation that there might be a connection between SQCD string and domain walls in the large $N$ limit. In particular, based
on the assumption that the SQCD domain walls are BPS saturated, Witten argued that in the large $N$ limit the domain walls connecting two vacua labeled by $k$ and $k^\prime=k+1$ appear to be objects that are not solitons from the SQCD string viewpoint but rather look like
D-branes \cite{Pol} on which the SQCD string should be able to end. Such D-brane-like domain walls have recently appeared in \cite{witt} in the context of the AdS/CFT correspondence \cite{ads}. 

{}It is desirable to verify the above assertions explicitly since they have important implications for SQCD. One of the key observations is that there exist D-branes on which the SQCD string can end \cite{witten}. In particular, the chromo-electric flux contained in the SQCD string that ends on the SQCD D-brane spreads along the D-brane, but away from the D-brane it is trapped
in the thin flux tube corresponding to the SQCD string \cite{witten,flux}. 
In fact, a test electric charge can only come off a D-brane (in the transverse direction) if it is connected to it by the SQCD string. Another way of putting this statement is that the D-branes should support excitations that behave as if they transform in the fundamental of $SU(N)$ albeit there are no such fields in the original 
super-Yang-Mills Lagrangian \cite{witten}. Moreover, naively one expects that the pure superglue SQCD contains only closed strings. However, if large $N$ domain walls indeed behave as D-branes, then we must also include open strings in the corresponding (non-critical) string theory to be able to describe these domain walls. This, in a sense, is analogous to what happens in Type II string theory: perturbatively we only expect closed strings; however, to capture non-perturbative dynamics we must also include open strings which end on the corresponding D-branes.   

{}The purpose of this paper is to give more evidence supporting the D-brane picture for the large $N$ SQCD domain walls. In particular, using an indirect construction, we obtain solutions for
the large $N$ SQCD domain walls which are BPS saturated. Also, it is important to check that the D-brane picture for the domain walls connecting two vacua labeled by $k$ and $k^\prime=k+j$ with $|j|=1$ is consistent with the fact that the domain walls connecting such vacua with $|j|\sim N$ are expected to behave as {\em solitons} of the SQCD string. Moreover, 
{\em a priori} it is far from being obvious that the domain walls connecting two vacua labeled by $k$ and $k^\prime=k+j$ with $|j|\sim N^\alpha$, $0<\alpha<1$,
are consistent with the effective SQCD string description. In particular, what is the interpretation of such domain walls in the context of the SQCD string?

{}It is, however, difficult to explicitly construct these domain walls (or even check that they are BPS saturated) directly in the context of SQCD: the order parameter $\lambda\lambda$ is a composite operator, and we are dealing with a strong coupling regime. 
Nonetheless, in this paper we use an approximating model which in a certain limit reduces to the large $N$ SQCD\footnote{More precisely, it appears to be in the same universality class as the large $N$ SQCD
meaning that the two theories are the same at low enough energies.}. In fact, using this approximating model we will be able to solve for the {\em exact} shapes of domain walls for $|j|=1$ (these domain walls are interpreted as SQCD D-branes), and also for $|j|\sim N$ (these domain walls are interpreted as solitons of the SQCD string). 
More precisely, we will obtain the corresponding exact solutions in a certain coordinate system which depends on the K{\"a}hler potential. The latter we will assume to be non-singular (in the regime we are interested in here).
Moreover, it will become clear that these solutions are BPS saturated. Also, we will argue that for all other values of $|j|$ the corresponding domain walls in the large $N$ limit
are described by $|j|$ D-branes separated    
by large distances. This provides an important check for the consistency of the D-brane interpretation and fits in the SQCD string picture.

{}The exact results we obtain in this paper are partly due to the fact that the domain wall problem reduces to a two dimensional problem, namely, of finding soliton configurations in two dimensional ${\cal N}=2$ massive quantum field theories. These theories have been studied
extensively (in fact, one of the original motivations to study them was precisely their relevance to domain walls in four dimensional ${\cal N}=1$ supersymmetric (gauge) theories), and the vast knowledge of their properties should not be overlooked when studying domain walls.

{}The rest of this paper is organized as follows. In section II we review some well known facts about ${\cal N}=1$ supersymmetric QCD. In section III we give a general discussion of domain walls in ${\cal N}=1$ supersymmetric theories. In section IV we construct the approximating model, and solve for the domain wall shapes in the large $N$ SQCD. In particular, we explicitly construct the BPS solutions corresponding to the D-branes and solitons. 
We also identify the limit in which the approximating model reduces to the large $N$ SQCD.
In section V we discuss the zero modes localized on the domain walls. In particular, we identify the zero modes localized on the D-branes (which are described by a three dimensional field theory) with the $U(1)$ gauge supermultiplet in three dimensions as required by the D-brane interpretation.
In section VI we discuss various issues regarding the large $N$ SQCD string and domain walls.   

\section{Preliminaries}\label{prelim}

{}In this section we review some well known facts about ${\cal N}=1$ supersymmetric QCD. In
particular, we briefly discuss gaugino condensation in SQCD without matter. We then
consider domain walls in SQCD. In the subsequent sections we will argue that
these domain walls are BPS saturated. 

\subsection{Gaugino Condensation in SQCD}

{}Consider ${\cal N}=1$ supersymmetric QCD with $SU(N)$ gauge group and no
matter. Let $\Lambda$ be the dynamically
generated scale. More precisely, if $g(\mu)\ll1$ is the Yang-Mill gauge coupling constant at some energy scale $\mu$, then $\Lambda=\mu\exp(-8\pi^2/b_0 g^2(\mu))$ is the scale at which the one-loop
renormalized gauge coupling blows up. (Here $b_0=3N$ is the one-loop $\beta$-function coefficient.)

{}In this theory there is a non-perturbative superpotential
due to gaugino condensation:
\begin{equation}\label{gaugino}
 {\cal W}=N\langle\lambda\lambda\rangle~.
\end{equation}    
The gaugino condensate is given by \cite{SV}
\begin{equation}\label{cond}
 \langle \lambda\lambda\rangle=a(N) \exp(2\pi i k/N) \Lambda^3~.
\end{equation}
Here $k=0,\dots,N-1$, that is, there are $N$ inequivalent vacua corresponding to $N$ different
phases for the gaugino condensate. This is due to spontaneous breaking of ${\bf Z}_N$ chiral symmetry. The factor $a(N)$ (which {\em a priori} can depend on $N$) is a positive number, and it depends on the subtraction scheme. However, 
the $N$ dependence of $a(N)$ becomes important in the large $N$ limit, in particular, 
consistency of the SQCD string interpretation for the large $N$ domain walls suggests 
that $a(N)\sim N$ \cite{witten}. This also naturally follows from the fact that the gaugino condensate $\langle\lambda\lambda\rangle$ involves a trace over the gauge indices, and the resulting vev should contain the corresponding second Casimir factor which in this case is simply $N$.

{}The superpotential in (\ref{gaugino}) can be derived as follows \cite{rev}.
Consider ${\cal N}=1$ supersymmetric gauge theory with $SU(N)$ gauge group and $N$ flavors $Q_i$, ${\widetilde Q}^{\bar j}$ ($i,{\bar j}=1,\dots,N)$. The gauge invariant degrees of freedom are mesons ${M_i}^{\bar j}=Q_i{\widetilde Q}^{\bar j}$,  and baryons $B=\epsilon^{i_1\dots i_N}
Q_{i_1}\cdots Q_{i_N}$ and ${\widetilde B}=\epsilon_{{\bar j}_1\dots {\bar j}_N}
{\widetilde Q}^{{\bar j}_1}\cdots {\widetilde Q}^{{\bar j}_N}$. The classical moduli space in this theory receives quantum corrections which can be accounted for via the following superpotential \cite{rev}\footnote{Here we note that this superpotential itself depends on the 
subtraction scheme. In particular, ${\widetilde \Lambda}$ is identified with the one-loop 
scale only up to a multiplicative constant that depends on the subtraction scheme. This constant
equals 1 in the ${\overline{\mbox{DR}}}$ scheme.}:
\begin{equation}\label{QM}
 {\cal W}=A\left(\det(M)-B{\widetilde B}-{\widetilde \Lambda}^{2N}\right)~.
\end{equation}   
Here ${\widetilde \Lambda}$ is the dynamically generated scale of the theory, and $A$ is a
Lagrange multiplier ($A{\widetilde \Lambda}^{2N}=W_a W_a$ is the ``glue-ball'' field). 
To obtain SQCD without matter, let us add a tree-level superpotential of the following form:
\begin{equation}
 {\cal W}_{\mbox{\small tree}}=m{\mbox{Tr}} (M)~.
\end{equation}
This superpotential corresponds to giving mass $m$ to all $N$ flavors. For $m\gg{\widetilde 
\Lambda}$, at energy scales below $m$ we can integrate the heavy flavors out which leaves
us with SQCD without matter. The superpotential now 
reads
\begin{equation}
 {\cal W}=N\exp(2\pi ik/N) m {\widetilde \Lambda}^2~.
\end{equation} 
The scale matching implies that 
\begin{equation}
 m {\widetilde \Lambda}^2=a(N) \Lambda^3~.
\end{equation}
That is, the scale matching depends on the threshold corrections (due to the heavy quarks
whose mass is of order $m$), and, therefore, on the subtraction scheme. 
Thus, we arrive at (\ref{gaugino}).

\subsection{Domain Walls in SQCD}

{}According to (\ref{gaugino}) and (\ref{cond}), there are $N$ inequivalent vacua in $SU(N)$ SQCD without matter. We therefore expect that there exist domain walls separating 
these different vacua \cite{wall}\footnote{For other related works, see, {\em e.g.}, \cite{rel}.}.
On dimensional grounds we expect the tension of such a wall
to be of order $\Lambda^3$. It is, however, possible to compute the tension exactly in terms of the gaugino condensate provided that they are BPS
saturated \cite{wall}\footnote{This was argued to be the case for $SU(2)$ in \cite{wall}.}. 
Consider the domain wall separating the
vacua with $\langle \lambda\lambda\rangle_k=a(N) \exp(2\pi i k/N) \Lambda^3$ and
$\langle \lambda\lambda\rangle_{k^\prime}=a(N) \exp(2\pi i k^\prime/N) \Lambda^3$.
The tension of the wall (provided that it is BPS saturated) is nothing but the value of the central
charge $Q_{kk^\prime}$ (in the corresponding central extension of the ${\cal N}=1$ superalgebra). The
central charge is proportional to (the absolute value of) the difference between the values
of the superpotential in the two vacua:
\begin{equation}
 Q_{kk^\prime}={1\over 8\pi^2} \vert {\cal W}_k-{\cal W}_{k^\prime}\vert~.
\end{equation}
We therefore have the following tension for the domain wall:
\begin{equation}
 T_{kk^\prime}={N\over 4\pi^2} a(N) \left| \sin\left({\pi(k-k^\prime)\over N}\right) \right|
 \Lambda^3~.
\end{equation}

{}It is, however, difficult to explicitly construct these domain walls (or even check that they are BPS saturated) directly in the context of SQCD: the order parameter $\lambda\lambda$ is a composite operator, and we are dealing with a strong coupling regime. Nonetheless, in the subsequent sections we will see that using an indirect construction ({\em i.e.}, by using a certain trick) it is possible to gain some insight into
the structure of these domain walls. In particular, we will argue that 
these domain walls are indeed BPS saturated. Moreover, in the large $N$ limit we will even be able to obtain the exact shapes of these domain walls. (More precisely, we will do so in a certain coordinate system which depends on the K{\"a}hler potential.)

{}The presence of BPS saturated domain walls in SQCD has important implications \cite{witten}. Strongly coupled SQCD in the large $N$ limit is believed to be described by a string theory with the string coupling $\lambda_s\sim 1/N$ \cite{thooft}. Extended solitons in this string theory are expected to have tension which goes as $1/\lambda^2_s\sim N^2$. It is not difficult to
see that not all of the above domain walls appear to be solitons from the SQCD string viewpoint. Thus, consider ``elementary'' domain walls (with $k^\prime=k+1$). In the large $N$ limit their tension goes as $a(N)$. For these objects to be SQCD solitons we  therefore would have to assume that $a(N)\sim N^2$. Let us now consider the walls with 
$\vert k^\prime -k\vert \sim N$. The wall tension in these cases goes as $Na(N)\sim N^3\sim
1/\lambda_s^3$ (provided that $a(N)\sim N^2$). This result is (at least) strange since
we do not expect such dependence on $\lambda_s$ for any BPS objects in the SQCD string context.

{}To avoid the above problem with the $\vert k^\prime -k\vert \sim N$ domain walls, we can
assume that $a(N)\sim N$. In this case these objects have tension that goes as $N^2$, and they
can be interpreted as solitons of the SQCD string. Now, however, we see that the ``elementary''
domain walls have tension which goes as $N\sim1/\lambda_s$. These cannot therefore be interpreted as solitons. In \cite{witten} it was suggested that such domain walls can be 
viewed as D-branes \cite{Pol} in the SQCD string context. Open SQCD strings then can end on these D-branes \cite{witten}.

{}Here we would like to point out another important property of the domain walls in SQCD. Thus, consider the domain walls with $|k^\prime-k|\sim N^\alpha$, where $0<\alpha<1$. If we assume that $a(N)\sim N$, then for these domain walls we naively get the tension that goes as $N^{1+\alpha}\sim 1/\lambda_s^{1+\alpha}$. Such states would be difficult to interpret in the context of the large $N$ SQCD string. In the following we will argue that in the large $N$ limit these domain walls actually correspond to $|k^\prime-k|$ ``elementary'' domain walls separated by large distances. This, in particular, avoids the problem with interpreting these domain walls 
in the context of the large $N$ SQCD string.

{}Before we conclude this section, we note that if we chose $a(N)$ to be independent of
$N$ we would run into a puzzle \cite{witten}: the $\vert k^\prime -k\vert \sim N$ domain walls would look like D-branes, but the ``elementary'' domain walls would have tension independent of
the string coupling (and, therefore, would have to be considered as fundamental, that is,
on the equal footing with the SQCD strings).

\section{Domain Walls in ${\cal N}=1$ Theories}

{}In this section we provide some details concerning generic properties of domain walls in ${\cal N}=1$ supersymmetric theories. In particular, we discuss 
the relation between the domain walls in four dimensional ${\cal N}=1$ theories and 
solitons in (massive) ${\cal N}=2$ quantum field theories in two dimensions.

{}Consider an ${\cal N}=1$ supersymmetric theory with one\footnote{Generalization of the following discussion to cases with multiple superfields is completely straightforward.} chiral superfield $X$. Let ${\cal W}(X)$ be the superpotential in this theory such that the
F-flatness condition ${\cal W}_X=0$ has a discrete set\footnote{For the sake of simplicity we will assume that these vacua are non-degenerate.} 
of solutions $X=X_a$, $a=1,\dots,N$.

{}In such a theory we expect presence of domain walls separating inequivalent vacua
$X=X_a$ and $X=X_b$, $a\not=b$.
More precisely, let $z$ be the spatial coordinate transverse to the wall. Then asymptotically
we have $X(z)\rightarrow X_a$ as $z\rightarrow-\infty$, and $X(z)\rightarrow X_b$ as $z\rightarrow+\infty$. Such domain walls may or may not be BPS saturated. 

{}Note that the problem at hand is really a two dimensional problem as the coordinates $x,y$
along the wall play no role in the discussion. Thus, our problem effectively reduces to that
of finding BPS solitons in two dimensional massive ${\cal N}=2$ quantum field theories. More
concretely, we are led to finding configurations of minimum energy (interpolating between
a pair of distinct vacua), which are solitonic solutions of the Landau-Ginsburg equations. 
For a general Landau-Ginsburg theory \cite{LG} the energy of a time-independent configuration is given by
\begin{equation}
 E_{ab}={1\over 2} \int_{-\infty}^{+\infty} dz \left(g\vert\partial_z X\vert^2 +V\right)~,
\end{equation}  
where $V=g^{-1}\vert W_X\vert^2$ is the scalar potential of the theory. 
Here $g(X,X^*)$ is the K{\"a}hler metric which is related to the K{\"a}hler potential ${\cal K}$ via
$g={\cal K}_{XX^*}$. 
Note that $E_{ab}$ can be rewritten as follows: 
\begin{equation}
 E_{ab}={1\over 2} \int_{-\infty}^{+\infty} dz g^{-1} \left| g\partial_z X^*-\exp(i\gamma)
 {\cal W}_X \right|^2 +{\mbox{Re}}\left(\exp(i\gamma)({\cal W}_{b}-{\cal W}_a)\right)~,
\end{equation}
where $\exp(i\gamma)$ is an arbitrary constant phase. Note that $E_{ab}$ is independent of $\gamma$. For the choice $\gamma=\gamma_{ba}$ we obtain the bound $E_{ab}\geq
|{\cal W}_{b}-{\cal W}_a|$ which is the BPS bound. Here
\begin{equation}
\exp(-i\gamma_{ba})= {{\cal W}_{b}-{\cal W}_a\over |{\cal W}_{b}-{\cal W}_a|}~.
\end{equation}
The BPS solutions (for which $E_{ab}= |{\cal W}_{b}-{\cal W}_a|$) are those that
satisfy the following equation:
\begin{equation}\label{BPS}
 g \partial_z X^* = \exp(i\gamma_{ba}) {\cal W}_X
\end{equation}
subject to the boundary conditions $X(z)\rightarrow X_{a,b}$ as $z\rightarrow \mp\infty$.
Going back to the domain walls in four dimensions, we have the exact same BPS equation
(\ref{BPS}), and the tension of the domain wall is given by
\begin{equation}
 T_{ab}={1\over 8\pi^2}|{\cal W}_{a}-{\cal W}_b|~.
\end{equation}

{}Typically, one does not know the exact form of the K{\"a}hler metric $g$. However, for the case of a single superfield $X$ we are considering here one can still make certain statements about the corresponding soliton solutions. Thus, let $X$ be a function of a new coordinate $z^\prime$ such that it satisfies the following equation 
\begin{equation}\label{BPS1}
 \partial_{z^\prime} X^* = \exp(i\gamma_{ba}) {\cal W}_X
\end{equation}
with the boundary conditions $X(z^\prime)\rightarrow X_{a,b}$ as $z^\prime \rightarrow \mp\infty$. Suppose we are able to find the corresponding solution for $X(z^\prime)$. Next, consider the following change of variables:
\begin{equation}\label{diff}
 \partial_{z^\prime} z(z^\prime) = g(X(z^\prime),X^* (z^\prime))~,
\end{equation}
where $X(z^\prime)$ is the corresponding solution. Note that if we express $X$ as a function of $z$, then $X(z)$ will satisfy the original BPS equation (\ref{BPS}) with the corresponding boundary conditions. The change of variables (\ref{diff}) is simply a diffeomorphism which is one-to-one as long as the K{\"a}hler metric $g$ is non-singular. Throughout this paper (when discussing domain walls in four dimensional ${\cal N}=1$ theories or the corresponding solitons in two dimensional ${\cal N}=2$ theories) we will assume that the corresponding K{\"a}hler metric $g$ is indeed non-singular. Moreover, we will always work in the coordinate system parametrized by $z^\prime$, but we will drop the prime for the sake of simplicity of the corresponding expressions. (This is effectively equivalent to the case where $g=1$.) Note that as long as the K{\"a}hler metric $g$ is non-singular,
the solution of (\ref{BPS1}) (with the appropriate boundary conditions) implies that the corresponding solution of (\ref{BPS}) exists. Moreover, this solution is BPS saturated.

{}The solitons in Landau-Ginsburg theories have been studied in detail \cite{LG,FMVW,CV}.
Many of these models are integrable. The general conditions for existence of BPS solutions were formulated in \cite{CV}. In this paper we will mainly be interested (for the reasons that will become clear in the next section) in the
so called $A_N$ models for which the superpotential is given by
\begin{equation}
 {\cal W}=X-{X^{N+1}\over{N+1}}~.
\end{equation} 
This model was studied in detail in \cite{FMVW}. In particular, there  are $N$ distinct vacua 
in this model with $X=X_k=\exp(2\pi ik/N)$, $k=0,\dots,N-1$. The soliton interpolating between the vacua
$X_k$ and $X_{k^\prime}$ has the energy
\begin{equation}\label{energy}
 E_{kk^\prime}={2N\over{N+1}}\left|\sin\left({\pi(k-k^\prime)\over N}\right)\right|~.
\end{equation}
All of these solitons are BPS saturated. (Here, as we have already pointed out, we take $g=1$.) Here we also mention that the (purely elastic) $S$-matrix
is exactly calculable in this model \cite{FMVW}.

\section{Large $N$ Domain Walls}

{}As we already mentioned, it is difficult to explicitly construct domain wall solutions in the
context of SQCD which is a strongly coupled theory. We can circumvent this difficulty by
considering a theory which is not SQCD itself but is closely related to it, in particular, it
shares the qualitative features of SQCD such as chiral symmetry breaking and presence of 
domain walls. More precisely, we would like to approximate SQCD by another theory which in an appropriate limit reproduces SQCD. We can expect that such an approximating theory
cannot reproduce (by taking a simple limit of any kind) SQCD at finite $N$: the latter is a highly
complicated model which we cannot hope to be able to solve in a simple manner. However,
as we discuss in this section, we can write down an approximating model which in a certain limit
reproduces the large $N$ SQCD. Using this trick, we, in particular, will be able to find the exact solutions for the shapes of the domain walls in the large $N$ SQCD.

\subsection{An Approximating Model}

{}As in section \ref{prelim}, let us consider ${\cal N}=1$ gauge theory with $SU(N)$ gauge group and $N$ flavors $Q_i$, ${\widetilde Q}^{\bar j}$ ($i,{\bar j}=1,\dots,N)$.
The non-perturbative superpotential in this theory is given by (\ref{QM}). 
Next, let us add a tree-level superpotential of the following form:
\begin{equation}
 {\cal W}_{\mbox{\small tree}}=X{\mbox{Tr}} (M)-{CN\over{N+1}} X^{N+1}~.
\end{equation}
Here $X$ is a chiral superfield neutral under the $SU(N)$ gauge group. Note that
the first term in this superpotential corresponds to giving mass $|X|$ to all $N$ flavors. 
For $|X|\gg{\widetilde \Lambda}$, at energy scales below $|X|$ we can integrate the heavy flavors out. This leaves us with SQCD without matter whose gaugino condensate is controlled
by the vev of $X$. Thus, the effective superpotential reads
\begin{equation}
 {\cal W}=N X {\widetilde \Lambda}^2-{CN\over{N+1}} X^{N+1}~.
\end{equation} 
The gaugino condensate is given by 
\begin{equation}
 \langle \lambda\lambda\rangle =X {\widetilde \Lambda}^2~.
\end{equation}
In other words, if $|X|$ is large compared with ${\widetilde \Lambda}$ everywhere, then
the gaugino condensate in this model is a linear function of $X$. By computing BPS
configurations in this model, we would therefore be effectively computing the spatial dependence of the gaugino condensate.

{}How is this model related to our starting point, namely, SQCD without matter? First, let us discuss the vacuum structure of this theory. The F-flatness condition ${\cal W}_X=0$
has $N$ solutions\footnote{There is also the moduli space at $X=0$. However, this will have no impact on the following discussions since all the configurations of interest here involve
$X\gg {\widetilde \Lambda}$.} (for definiteness here we choose $C$ to be positive): 
\begin{equation}
 \langle \lambda\lambda\rangle_k =X_k {\widetilde \Lambda}^2~,~~~
 \eta\equiv({\widetilde \Lambda}^{2-N}/C)^{1/N}~,
\end{equation}
where $X_k=\eta{\widetilde\Lambda}\exp(2\pi ik/N)$. Thus, this model has chiral
symmetry breaking, namely, the ${\bf Z}_N$ subgroup of the $R$-parity group is spontaneously
broken. There are domain walls between different vacua. These domain walls are BPS saturated.
This follows from our discussion in the previous section, and the fact that this model is nothing
but the $A_N$ model when viewed from the two dimensional ${\cal N}=2$ quantum field theory
viewpoint. 
Thus, this approximating model resembles SQCD. However, there are also differences. First, the theory contains heavy quarks with masses of order $\eta{\widetilde \Lambda}$. In order to be able to approach SQCD, we therefore must take the limit $\eta\rightarrow\infty$. Naively, this might appear to be enough to map this model to SQCD. This is, however, not the case. To see this, consider the value of the superpotential for a given
vacuum: 
\begin{equation}
 {\cal W}_k={N^2\over{N+1}}X_k {\widetilde \Lambda}^2=
 {N^2\over{N+1}} \langle \lambda\lambda\rangle_k ~.
\end{equation}   
Thus, the relation which would be correct in SQCD, namely, 
\begin{equation}
 {\cal W}_k=
 N \langle \lambda\lambda\rangle_k~, 
\end{equation}
is reproduced only in the limit $N\rightarrow\infty$. (This is required to get the same central charges as in SQCD.) So we must also take 't Hooft's large $N$
limit. In fact, we will see in the following that the two limits, namely, $\eta\rightarrow\infty$ and
$N\rightarrow\infty$, are not independent. The precise relation between these limits (that is,
which combination of $\eta$ and $N$ needs to be kept fixed) will become clear in
subsection D. Here we note that in this limit the mass of the field $X$ in any given vacuum $X=X_k$ is infinitely large, so that this model indeed reproduces large $N$ SQCD. However, for the domain wall solution interpolating between two distinct vacua $X=X_k$ and $X=X_{k^\prime}$ there is a zero mode of $X$ localized on the wall. In fact, this zero mode appears to be needed for interpreting the corresponding large $N$ domain walls as D-branes for the SQCD string. 

\subsection{``Elementary''  Domain Walls at Large $N$}

{}Next, consider the BPS equation (\ref{BPS1}) for the above approximating model. (Here and in the following we drop the prime and use $z$ instead of $z^\prime$.) In particular,
let us study it for the domain wall where at $z\rightarrow-\infty$ we have $X=X_k$, and
at $z\rightarrow+\infty$ we have $X=X_{k+1}$. We have been referring to this wall as an
``elementary'' wall. First, note that $\exp(-i\gamma_{k+1,k})=i\exp(2\pi ik/N+\pi i/N)$. Let 
$X=X_k Y$. Then we can rewrite (\ref{BPS}) in terms of $Y$:
\begin{equation}\label{Y} 
 \partial_z Y^*=-i\exp(-\pi i/N){N{\widetilde \Lambda} \over \eta}\left[1-Y^N\right]~.
\end{equation}   
The boundary conditions on $Y$ read: 
\begin{equation} 
 Y(z\rightarrow-\infty)=1~,~~~Y(z\rightarrow+\infty)=\exp(2\pi i/N)~.
\end{equation}
Thus, in the large $N$ limit ${\mbox{arg}}(Y)$ changes by a small amount, namely, from $0$
to $2\pi/N$. It is convenient to parametrize $Y$ as follows:
\begin{equation} 
 Y=(1-\rho/N)\exp(i(\phi+\pi)/N)~.
\end{equation}
Here $\rho$ is real. The boundary conditions then read:
\begin{equation} 
 \rho(z\rightarrow\pm\infty)=0~,~~~\phi(z\rightarrow\pm\infty)=\pm\pi~.
\end{equation}
Then (\ref{Y}) becomes a system of the following first order differential equations: 
\begin{eqnarray}
 &&\partial_z \phi=\Gamma \left[1+\exp(-\rho)\cos(\phi)\right]~,\\
 &&\partial_z \rho=-\Gamma \exp(-\rho)\sin(\phi)~.
\end{eqnarray}
Here we are taking the large $N$ limit and only keeping the leading terms. 
In particular, we have taken into account that $(1-\rho/N)^N\rightarrow\exp(-\rho)$
as $N\rightarrow\infty$. Also, we have introduced
\begin{equation} 
 \Gamma=N^2{\widetilde\Lambda}/\eta~.
\end{equation}

{}The above system of differential equations can be
integrated. Note that these equations do not explicitly contain $z$. This implies that
if $\phi(z)$ and $\rho(z)$ give a solution with the appropriate boundary conditions,
so will be $\phi(z+z_0)$ and $\rho(z+z_0)$ for any constant $z_0$. (This is simply the statement that the system possesses translational invariance in the $z$ direction.) There is a solution that
has a symmetry with respect to the reflection $z\rightarrow-z$. For definiteness, let us
focus on this solution. It is given by
\begin{eqnarray} 
 &&\cos(\phi)=(\rho-1)\exp(\rho)~,\\
 &&{\cal F}(\rho)=\Gamma |z|~.
\end{eqnarray}
Here
\begin{equation} 
 {\cal F}(\rho)=\int_{\rho}^{\rho_0} d\xi \left[\exp(-2\xi)-(1-\xi)^2\right]^{-{1\over 2}}~,
\end{equation} 
and $\rho_0=\rho(z=0) (\approx 1.278)$ is the solution of the following equation:
\begin{equation} 
 (\rho_0-1)\exp(\rho_0)=1~.
\end{equation}
For illustrative purposes we have plotted the shapes of $\rho(z)$ and $\phi(z)$ corresponding to the above exact solution in Fig.1 and Fig.2, respectively. Here we note that the ``width'' of the ``elementary'' walls is of order $\Gamma^{-1}$.

\subsection{Other Domain Walls at Large $N$}

{}Next, let us analyze the domain walls with $k^\prime-k=j$, where $j$ is a fixed integer, that is,
$|j|/N\rightarrow 0$ as $N\rightarrow\infty$. More precisely, we will first consider the cases where $|j|\sim 1$. In these cases it is convenient to parametrize $X$ as follows. Let $X=X_kY$ where $Y=(1-\rho/N)\exp(i(\phi+\pi j)/N)$. Then we have the following boundary conditions
($\rho$ is real): $\rho(z\rightarrow\pm\infty)=0$, $\phi(z\rightarrow\pm\infty)=\pm \pi j$. In the large $N$ limit the BPS equation (\ref{BPS1}) becomes a system of the following first order differential equations: 
\begin{eqnarray}
 &&\partial_z \phi=\Gamma \left[1-(-1)^j \exp(-\rho)\cos(\phi)\right]~,\\
 &&\partial_z \rho=(-1)^j \Gamma \exp(-\rho)\sin(\phi)~.
\end{eqnarray} 
It is not difficult to show that for $|j|\not=1$ the only non-trivial solution to this system of 
equations is the one corresponding to $|j|$ ``elementary'' domain walls separated by large distances (that is, distances much greater than the width $\Gamma^{-1}$ of the ``elementary'' walls). 

{}Let us understand this fact a bit better. In particular, at first it might appear strange that for $|j|\not=1$ we do not have any non-trivial solutions other than $|j|$ ``elementary'' domain walls placed at large distances from each other\footnote{Such a configuration strictly speaking is not a solution. More precisely, for $|j|\not=1$ there are no solutions. However, $|j|$ ``elementary'' domain walls separated by distances much greater than $\Gamma^{-1}$ are exponentially close 
to being a solution of the BPS equation.}. Thus, from the two dimensional quantum field theory viewpoint we expect to find non-trivial solutions for all values of $j$. However, at large $N$ there are certain simplifications. In particular, consider the energy $E_{k,k+j}$ of the corresponding solitons in the two dimensional quantum field theory. For $|j|/N\rightarrow 0$ as $N\rightarrow\infty$ we have $E_{k,k+j}=2\pi |j|/N =|j| E_{k,k+1}$. This implies that at large $N$
the corresponding solitons fall apart into $|j|$ ``elementary'' solitons\footnote{More precisely, for finite $N$ the solutions with $|j|\not=1$ can be viewed as ``bound states'' of ``elementary'' solitons with $|j|=1$ which become infinitely separated as $N\rightarrow\infty$.}. In the four dimensional language, at large $N$ the domain walls with $k^\prime=k+j$, $|j|\not=1$, fall apart into $|j|$ ``elementary'' domain walls. In the following we argue that this is consistent with the interpretation of the ``elementary'' domain walls as D-branes for the SQCD string. On the other hand, for finite $N$ we have $E_{k,k+j}<|j| E_{k,k+1}$. This is, however, a subleading effect in the large $N$ limit. We therefore conclude that the domain walls with $|j|\not=1$ are stable (in the sense that they do not fall apart into ``elementary'' domain walls) only at finite $N$. 

{}To address the above issue in more detail, let us consider the domain walls with $k^\prime-k=j$ more carefully. Let us parametrize $X$ as $X=X_kY$. Then the boundary conditions on $Y$ read: $Y(z\rightarrow-\infty)=1$, $Y(z\rightarrow+\infty)=\exp(2\pi ij/N)$. The BPS equation 
(\ref{BPS1}) then reads:
\begin{equation}\label{BPSY}
 \partial_z Y^*=-i \exp(-\pi i j/N)\Delta \left[1-Y^N\right]~.
\end{equation} 
Here we have introduced
\begin{equation}
 \Delta=N{\widetilde \Lambda}/\eta~.
\end{equation}
Note that $\Gamma=N \Delta$.

{}Suppose there is a BPS solution such that $0<|Y|<1$ for 
$z$ taking values in some finite interval.
By this me mean that
$Y^N\rightarrow 0$ as $N\rightarrow\infty$ for these values of $z$. Thus, we can try the following ans{\"a}tz  (whose self-consistency is to be verified) for solving the BPS equation (\ref{BPS}):
\begin{equation}
 \partial_z Y^*=-i \exp(-\pi i j/N)\Delta~,~~~z_- \leq z\leq z_+~,
\end{equation} 
where it is assumed that $Y(z)=1$ for $z<z_-$, and $Y(z)=\exp(2\pi i j/N)$ for $z>z_+$.
The solution to the above ans{\"a}tz is given by
\begin{eqnarray}
 &&Y(z)=1~,~~~z<z_-~,\\
 &&Y(z)=1+i\Delta\exp(i\pi j/N) (z-z_- )~,~~~z_- \leq z\leq z_+ ~,\\
 &&Y(z)=\exp(2\pi i j/N)~,~~~z>z_+~,
\end{eqnarray}
where $z_+=z_-+2L$, and 
\begin{equation}\label{L}
 L=\Delta^{-1}\sin\left(\pi j/N \right)~.
\end{equation}
Here $z_-$ is arbitrary.

{}Note that there are constraints for the validity of the above solution. First, we must consider the ``edge'' effects. Namely, for $z=z_- +\epsilon$, and $z=z_+-\epsilon$, where $\epsilon\ll L$, we
have $1-|Y(z)|\approx(\epsilon/L) \sin^2(\pi j/N)$. Now suppose that 
as $N\rightarrow\infty$ we have $|j|\sim N^\alpha$, where $0<\alpha<1$. Then the assumption $Y^N\rightarrow 0$ as $N\rightarrow\infty$ breaks down at 
distances $\epsilon\sim (L/N) N^{2(1-\alpha)} \gg (L/N)=\Gamma^{-1}$. Now recall that $\Gamma^{-1}$ is the width of the ``elementary'' domain walls. Thus, for $|j|\sim N^\alpha$, $0<\alpha<1$, the above solutions would break down due to the ``edge'' effects at distances (from the ``edge'') which are much greater than the width of the ``elementary'' domain walls. 
This implies that the correct description of the domain walls with $|j|\sim N^\alpha$, $0<\alpha<1$, appears to be in terms of $|j|$ ``elementary'' domain walls (separated by large distances) rather than as solitons of the large $N$ SQCD string. 

{}Next, consider the domain walls with $|j|\sim N$ in the large $N$ limit. For such domain walls the ``edge'' effects become important at distances $\epsilon \sim L/N=\Gamma^{-1}$. Thus, the above solution for such domain walls can be trusted for all $z$ except for the small regions near the ``edges''. The size of these regions is of order $\Gamma^{-1}$, that is, the width of the ``elementary'' domain wall. This implies that the domain walls with $|j|\sim N$ can be viewed as solitons of the SQCD string provided that $L$ is finite and, therefore, the width of the ``elementary'' domain walls ({\em i.e.}, the width of the D-branes) is vanishing in the large $N$ limit. 

{}Before we end this subsection, we point out the second restriction on the validity of the
above solution for the $|j|\sim N$ domain walls interpreted as solitons of the SQCD string. The following restriction has to do with the fact the approximating model we have been using is strictly speaking valid 
only if $|X|{\ \lower-1.2pt\vbox{\hbox{\rlap{$>$}\lower5pt\vbox{\hbox{$\sim$}}}}\ }{\widetilde \Lambda}$. In the above solution we have $|X|=\eta{\widetilde \Lambda}|Y|$. On the other hand, it is not difficult to see that 
\begin{equation}
 {\mbox{min}}(|Y(z)|)=|Y(z_- +L)|=\left|\cos\left({\pi j\over N}\right)\right|~.
\end{equation}
This expression vanishes for $j=N/2$ (which is possible for even $N$), and can be arbitrarily small for $j=(N+j^\prime)/2$ in the large $N$ limit. It is not difficult to see that under these circumstances $|X(z)|{\ \lower-1.2pt\vbox{\hbox{\rlap{$<$}\lower5pt\vbox{\hbox{$\sim$}}}}\ }{\widetilde\Lambda}$ for $z=z_-+L+\epsilon$, where $|\epsilon|{\ \lower-1.2pt\vbox{\hbox{\rlap{$<$}\lower5pt\vbox{\hbox{$\sim$}}}}\ } L/\eta$. In the next
subsection we will see that in the appropriate limit 
$L/\eta\ll \Gamma^{-1}$. This implies that the above ``soliton'' solutions 
for $|j|\sim N$ can be trusted everywhere in the large $N$ limit up to the ``edge'' 
effects due to the D-branes, and (possible) ``core'' effects due to the quarks $Q_i,{\widetilde Q}^{\bar j}$ becoming too light. (These effects, however, are negligible as they affect the solution only in infinitesimally small regions.)  

\subsection{Summary}

{}The above discussion leads us to the following picture. The width of the soliton solutions with
$|j|\sim N$ is of order $L\sim \Delta^{-1}$. In SQCD we expect this to be of order $\Lambda^{-1}$, where $\Lambda$ is the dynamically generated scale of SQCD. Note that this is also consistent with the fact that $L\sim L_s$, where $L_s\sim \Lambda^{-1}$ is the SQCD string length.
(The SQCD string tension is expected to be $T_s\sim \Lambda^2$.) Then the width of the
``elementary'' domain walls, interpreted as D-branes for the (open) SQCD strings, is of order
$\Gamma^{-1}=\Delta^{-1}/N\sim L/N$, which is vanishing for finite $L$. This is consistent with the (at least naive) expectation that in a weakly coupled string theory\footnote{The SQCD string theory does {\em not} appear to be a critical string theory. The following statement is correct for a critical string theory but is {\em a priori} far from being obvious for non-critical strings.} D-branes have vanishing width.

{}Another important point is that the domain walls with $|j|\not=1$ appear to fall apart into 
$|j|$ D-branes (separated by large distances) unless $|j|/N$ is a finite number as $N\rightarrow \infty$. This is important for two reasons. Thus, had we found additional domain walls with $|j|\sim N^\alpha$, $0<\alpha<1$, we would definitely have a puzzle since the tension for these domain walls would go as $1/\lambda_s^{1+\alpha}$. Second, it appears that we {\em cannot} stack D-branes on top of each other. This is consistent with the rest of the above picture, since had we been able to do so, we would have to have a moduli space associated with the positions of D-branes. These would have to be described by adjoint scalars in the effective field theory. However, we do not expect any
such adjoint scalars in this model due to ${\cal N}=1$ supersymmetry in $2+1$ dimensions (in the world-volumes of the D-branes - see the next section). So the above conclusions seem to be consistent with this fact as well. 

{}We are now ready to determine the precise limit that we ought to take to
recover (the large $N$) SQCD consistent with the above discussions. First, we conclude that
we must have $\Delta\sim\Lambda$. On the other hand, the gaugino condensate $|\langle\lambda\lambda
\rangle_k|=\eta{\widetilde \Lambda}^3$. According to our discussion in section \ref{prelim},
we must have $|\langle\lambda\lambda\rangle_k|\sim N\Lambda^3$. Thus, we have two
relations:
\begin{eqnarray} 
 &&N{\widetilde\Lambda}/\eta\sim\Lambda~,\\
 &&\eta{\widetilde \Lambda}^3\sim N\Lambda^3~.
\end{eqnarray}
This implies that we must take the following limit:
\begin{equation} 
 \eta,N\rightarrow\infty~,~~~\eta/N\equiv\zeta^3={\mbox{fixed}}~,~~~\Lambda/{\widetilde  
 \Lambda}\sim\zeta={\mbox{fixed}}~.
\end{equation}
Here $\zeta$ is a finite number. (Note that $\Lambda$ is fixed to a finite value.) However, we must actually have $\zeta\gg 1$ 
(that is, $\Lambda\gg {\widetilde \Lambda}$) so that the procedure of integrating out the heavy quarks $Q_i,{\widetilde Q}^{\bar j}$ is valid.  
We note that in this limit the above model is a valid description as far as the ``elementary'' domain walls are concerned, that is, none of the
assumptions (such as, say, validity of integrating out the heavy quarks) on which it was built
break down. In particular, note that the change in the absolute value of $X$ through the walls
corresponding to the D-branes is subleading in the large $N$ limit. Thus, the inequality $|X|\gg{\widetilde \Lambda}$ always holds. Also, the masses of the heavy quarks $Q_i,{\widetilde Q}^{\bar j}$ (which we have integrated out) are of order $m\sim |X|\sim \eta{\widetilde \Lambda}$. On the other hand, the width of the D-branes is of order $\Gamma^{-1}$, and we have $m/\Gamma\sim(N/\eta)^2=\zeta^{-6}\ll 1$. This implies that integrating out these heavy quarks is completely legitimate.

\section{Zero Modes}

{}The above picture appears to be consistent with the arguments of \cite{witten} as well as various discussions in the previous sections. However, if we are to interpret some of the large $N$ domain walls as D-branes, we expect (at least naively) that there should be a $U(1)$ gauge field living in the world-volume of such a D-brane \cite{Pol}. More precisely, we expect a three dimensional ${\cal N}=1$ $U(1)$ gauge supermultiplet localized on the D-brane. Can we identify such a supermultiplet in the above picture?

{}This is where the zero modes of the chiral superfield $X$ in the above approximating model find a natural interpretation. Note that in the above limit the mass of the field $X$ in any given vacuum $X=X_k$ is infinitely large. This is consistent with the interpretation of the approximating model as reducing to the large $N$ SQCD (where in any given vacuum we do not expect additional singlet fields). However, for the solution describing a domain wall separating two distinct vacua $X=X_k$ and $X=X_{k^\prime}$ there always exist zero modes of $X$ localized on the domain wall corresponding to the Goldstone modes of broken translational invariance and supersymmetry. More precisely, since $X$ in the four dimensional ${\cal N}=1$ supersymmetric theory is a chiral superfield containing a (two-component) chiral fermion and a complex boson, the corresponding zero modes (corresponding to a BPS solution that breaks half of the supersymmetries) consist of a one-component fermion and a real scalar in three dimensions. These form an ${\cal N}=1$ ``chiral'' supermultiplet in three dimensions. However, in three dimensions this is dual to a $U(1)$ vector supermultiplet consisting of a vector boson (with one physical degree of freedom) plus its ${\cal N}=1$ superpartner. This is due to the fact that a $U(1)$ vector boson in three dimensions can be dualized into a real scalar. Thus, for the domain walls corresponding to the D-branes the zero modes of $X$ are interpreted (in the dual ``magnetic'' picture) as the $U(1)$ gauge supermultiplet living in the ($2+1$ dimensional) world-volume of the D-brane.

{}Let us discuss the zero modes localized on a D-brane in a bit more detail. For simplicity, let us confine our attention to the bosonic zero mode. (The fermionic zero mode can be obtained by the corresponding supersymmetry transformation.) Thus, consider the domain wall separating two vacua $X=X_k$ and $X=X_{k+1}$. Let $X(z)$ be the corresponding BPS solution. Then the zero mode is given by:   
\begin{equation} 
 \Phi(z,x_\mu)=\partial_z X(z) \phi (x_\mu)~.
\end{equation}
Here $\mu=0,1,2$. 

{}It is not difficult to see that for this D-brane solution the zero mode is localized in the region of the size $\Gamma^{-1}$ which is the same as the width of the D-brane. (Recall that in the limit we are taking here this width is vanishingly small.) Then according to the above discussion the bosonic and fermionic zero modes correspond to the three dimensional ${\cal N}=1$ $U(1)$ gauge supermultiplet localized in the world-volume of the D-brane.  

{}Here the following remarks are in order. First, note that the above zero modes correspond to the translations in the direction transverse to the D-brane. On the other hand, the change in the absolute value of $X$ through the corresponding domain wall is subleading in the large $N$ limit. Thus, the zero mode is related to the change in the phase of $X$. The latter is periodic, so
that the dualization procedure (to a $U(1)$ gauge multiplet) leads to quantization of the corresponding $U(1)$ charge\footnote{We would like to thank Juan Maldacena for a discussion on this point.}.   

{}Second, at low energies we expect a {\em free} $U(1)$ gauge multiplet localized on the D-brane. This requires that the dual scalar also be free. This is guaranteed by the fact that here we have a Goldstone mode (of broken translational invariance in the $z$-direction) with purely derivative interactions suppressed by the wall tension. (The corresponding interactions involving the fermionic component follow by the ${\cal N}=1$ supersymmetry transformation in three dimensions.) In particular, at low energies the theory is indeed free. 

{}Thus, consider the low energy effective field theory corresponding to the zero modes of $X$ localized on the D-brane. In particular, let us discuss the bosonic part of the corresponding low energy Lagrangian, namely, let us only keep the terms arising through the induced metric 
$G_{\mu\nu}$ on the wall: 
\begin{equation}\label{el}
 {\cal L}\sim \sqrt{G}={\det}^{1/2} \left(\eta_{\mu\nu}+\partial_\mu \phi
\partial_\nu \phi\right)~,
\end{equation}
where $G=\det(G_{\mu\nu})$. (Here $\eta_{\mu\nu}$ is the three dimensional Minkowskian metric, and we have normalized the field $\phi$ so that $\partial_\mu \phi$ is dimensionless.) On the other hand, we have the following vector-scalar duality transformation in $2+1$ dimensions ($\mu,\nu,\lambda=0,1,2$):
\begin{equation}\label{dual} 
 F_{\mu\nu}=\epsilon_{\mu\nu\lambda}\partial^\lambda\phi~.
\end{equation}
Here $F_{\mu\nu}=\partial_\mu A_\nu - \partial_\nu A_\mu$ is the $U(1)$ field strength, $A^\mu$ is the vector potential, and $\phi$ is the dual scalar. The Lagrangian in (\ref{el}) can be rewritten as  
\begin{equation}
 {\cal L}\sim {\det}^{1/2} \left(\eta_{\mu\nu}+F_{\mu\nu}\right)~.
\end{equation}
This is nothing but the Born-Infeld Lagrangian for a $U(1)$ gauge field (as expected for the low energy effective Lagrangian of a D-brane). At low energies it reduces to the free 
Maxwell theory in $2+1$ dimensions:  
\begin{equation}
 {\cal L}= -{1\over 4g_D^2} {\cal F}_{\mu\nu} {\cal F}^{\mu\nu}~.
\end{equation}
Here ${\cal F}_{\mu\nu}\equiv \Lambda^2 F_{\mu\nu}$ has conventional dimension of mass squared (whereas $F_{\mu\nu}$ is dimensionless). The $U(1)$ gauge coupling is given
by:
\begin{equation}
 g_D^{-2}=T_D/\Lambda^4~,
\end{equation}
where $T_D$ is the D-brane tension.

\section{Discussion}

{}Let us briefly summarize the main points in the previous discussions. We consider an approximating model which in a particular limit reduces to the large $N$ SQCD. This model is constructed by starting from ${\cal N}=1$ $SU(N)$ gauge theory with $N$ flavors
plus a gauge singlet. The tree-level superpotential corresponds to the chiral ${\bf Z}_N$ symmetry breaking with all the flavors acquiring large masses. There are $N$ inequivalent vacua in this model. In a particular limit where the vev of the gauge singlet is large, and $N\rightarrow\infty$, this theory reduces to the large $N$ SQCD without matter in which we have
gaugino condensate. The inequivalent vacua then correspond to different phases $\exp(2\pi i k/N)$, $k=0,\dots, N-1$ of  the gaugino condensate.

{}Using our approximating model, we then are able to solve for the exact shapes of the
domain walls (separating the vacua labeled by $k$ and $k^\prime=k+j$, $j=1,\dots, N-1$)
in the large $N$ SQCD. 
These exact solutions are obtained in a certain coordinate system which depends on the K{\"a}hler metric (which we assume to be non-singular).
In particular, we find that there are two basic types of solutions.
The first type corresponds to the ``elementary'' domain walls with $|j|=1$. These are interpreted as D-branes of the corresponding large $N$ SQCD string. The tension of these domain walls
goes as $N\sim1/\lambda_s$ (with the appropriate choice of the subtraction scheme dictated by
the requirement that all domain walls have proper interpretation in the SQCD string context, which, in particular, implies that the gaugino condensate $|\langle\lambda\lambda\rangle_k|\sim N\Lambda^3$). The second type of domain walls occur for $|j|\sim N$. These are interpreted
as solitons of the corresponding large $N$ SQCD string. The tension of these domain walls
goes as $N^2\sim1/\lambda_s^2$. For other values of $|j|$ the corresponding domain walls (in the large $N$ limit) appear to be described as configurations composed of $|j|$ ``elementary'' domain walls (D-branes) separated by distances large compared with the D-brane width. (The
latter is of order $\Lambda^{-1}/N\rightarrow 0$. In contrast, the solitons have finite width of order
$\Lambda^{-1}$.)

{}We have also argued that the zero mode (corresponding to the singlet $X$) localized in the world-volume of a D-brane is dual to the ${\cal N}=1$ $U(1)$ vector supermultiplet in $2+1$ dimensions. Open SQCD strings should be able to end on these D-branes. Then, these 
D-branes support excitations that behave as if they transform in the fundamental of $SU(N)$ \cite{witten}. In other words, 
as far as the quantum numbers are concerned, 
the region where the SQCD string ({\em i.e.}, the chromo-electric flux tube)
enters the D-brane looks like a state in the fundamental of $SU(N)$. 
This simply follows from the flux conservation.      

{}The above discussions appear to imply that whatever the SQCD string theory is, it must contain D-branes and open strings as well as (expected) closed strings. Here one can draw an analogy with Type II string theory which perturbatively contains only the closed string states, but must be augmented with D-branes and open string sectors to describe non-perturbative phenomena. In fact, in the SQCD string theory it seems to be necessary to include the open strings and D-branes (in the large $N$ limit). These states are {\em not} solitons of the effective closed string theory, but must be introduced or else phenomena such as chiral symmetry breaking and appearance of (some of) the domain walls cannot be described. At present it is unclear what this SQCD string theory is. However, we do expect that it is a non-critical string theory with varying string tension \cite{polyakov}. It would be interesting to understand the role of D-branes in such string theories. This might shed additional light on the structure of SQCD. 

{}Finally, we would like to make the following comment. The particular approximating model we used in this paper for our discussion of domain walls in the large $N$ SQCD is not the only possible starting point. Thus, one can equally successfully start from, say, ${\cal N}=1$ $SU(N)$
gauge theory with one flavor with a large mass. Upon integrating out this heavy flavor we obtain SQCD. Then the domain walls can be studied much in the same way as we did in this paper. (In fact, this was the setup used in \cite{wall} to study domain walls in the case of $SU(2)$.)
The resulting domain walls (for large $N$) are BPS saturated, and their shapes in the coordinate system where the corresponding K{\"a}hler metric is eliminated are related to those obtained in this paper by an appropriate diffeomorphism accompanied by a (non-linear) change of variables
(which changes the K{\"a}hler potential). The final answer is, however, the same. One should also be able to start from the    
effective Veneziano-Yankielowicz superpotential \cite{VY}. In this case the corresponding two dimensional ${\cal N}=2$ quantum field theory reduces to the ${\bf CP}^{N-1}$ model. (See, {\em e.g.}, \cite{CV} for a discussion of this model.) Generically, (unless the ``glue-ball'' field becomes too small) this description should also be adequate. Here we should mention that there is a technical problem with using the effective Veneziano-Yankielowicz superpotential, namely, a logarithmic branch cut appearing in the superpotential. It would be interesting to see whether one can obtain domain wall shapes in this approach and map them to those discussed in this paper via the   
appropriate diffeomorphism plus change of variables.

\acknowledgements

{}We would like to thank Philip Argyres, Michael Bershadsky, Edi Gava, Andrei Johansen, 
Juan Maldacena,
K.S. Narain, Pran Nath, Michael Shifman, Tom Taylor, Henry Tye, Cumrun Vafa and Piljin Yi
for discussions. The work of Z.K. was supported in part by the grant NSF PHY-96-02074, 
and the DOE 1994 OJI award. Z.K. would also like to thank Albert and Ribena Yu for 
financial support.

\newpage
\begin{figure}[t]
\hspace*{2.6 cm}
\epsfxsize=10 cm
\epsfbox{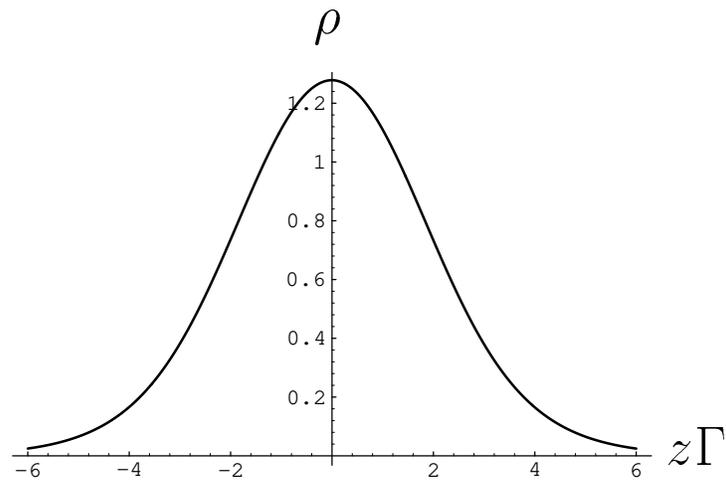}
\caption{The shape of $\rho(z)$. }
\end{figure}

\bigskip

\begin{figure}[t]
\hspace*{2.6 cm}
\epsfxsize=10 cm
\epsfbox{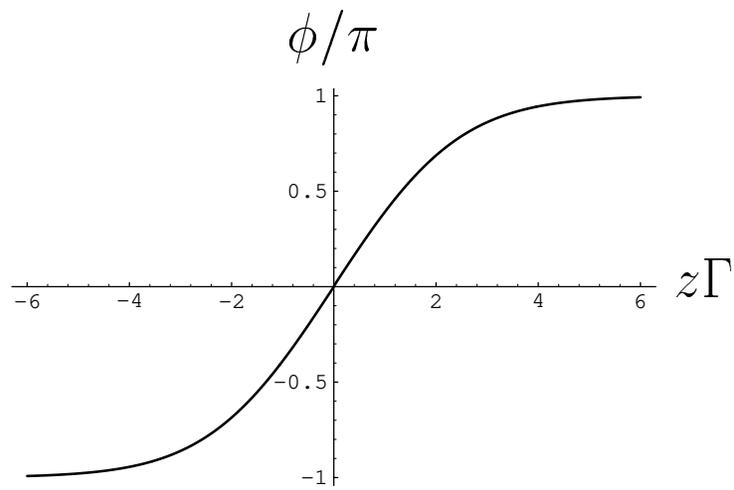}
\caption{The shape of $\phi(z)$. }
\end{figure}

\end{document}